\documentclass[11pt]{article}

\usepackage[final]{acl}

\usepackage{times}
\usepackage{latexsym}
\usepackage{booktabs}
\usepackage{amssymb}
\usepackage{pifont}
\usepackage{makecell} 
\newcommand{\cmark}{\checkmark}
\newcommand{\xmark}{\ding{55}}
\usepackage{colortbl} 
\usepackage{multirow}
\usepackage{arydshln}
\usepackage{threeparttable}
\usepackage[minted, most]{tcolorbox}

\usepackage[T1]{fontenc}

\usepackage[utf8]{inputenc}

\usepackage{microtype}

\usepackage{inconsolata}

\usepackage{graphicx}

\title{WenetSpeech-Wu: Datasets, Benchmarks, and Models for a Unified Chinese Wu Dialect Speech Processing Ecosystem}

\author{
 \textbf{Chengyou Wang\textsuperscript{1}\thanks{Equal Contribution. $\dagger$ Corresponding Author.}},
 \textbf{Mingchen Shao\textsuperscript{1$\ast$}},
 \textbf{Jingbin Hu\textsuperscript{1$\ast$}},
 \textbf{Zeyu Zhu\textsuperscript{1$\ast$}},
 \textbf{Hongfei Xue\textsuperscript{1}}, 
 \\
 \textbf{Bingshen Mu\textsuperscript{1}},
 \textbf{Xin Xu\textsuperscript{2}},
 \textbf{Xingyi Duan\textsuperscript{6}},
 \textbf{Binbin Zhang\textsuperscript{3}},
 \textbf{Pengcheng Zhu\textsuperscript{3}}, 
 \\
 \textbf{Chuang Ding\textsuperscript{4}},
 \textbf{Xiaojun Zhang\textsuperscript{5}},
 \textbf{Hui Bu\textsuperscript{2}},
 \textbf{Lei Xie\textsuperscript{1$\dagger$}},
\\
 \textsuperscript{1}Audio, Speech and Language Processing Group (ASLP@NPU),\\ Northwestern Polytechnical University\\
 \textsuperscript{2}Beijing AISHELL Technology Co., Ltd.,
 \textsuperscript{3}WeNet Open Source Community\\
 \textsuperscript{4}Moonstep AI,
 \textsuperscript{5}Xi'an Jiaotong-Liverpool University,
 \textsuperscript{6}YK Pao School
}

\begin{document}
\maketitle
\begin{abstract}
Speech processing for low-resource dialects remains a fundamental challenge in developing inclusive and robust speech technologies.
Despite its linguistic significance and large speaker population, the Wu dialect of Chinese has long been hindered by the lack of large-scale speech data, standardized evaluation benchmarks, and publicly available models.
In this work, we present \textbf{WenetSpeech-Wu}, the first large-scale, multi-dimensionally annotated open-source speech corpus for the Wu dialect, comprising approximately 8,000 hours of diverse speech data.
Building upon this dataset, we introduce \textbf{WenetSpeech-Wu-Bench}, the first standardized and publicly accessible benchmark for systematic evaluation of Wu dialect speech processing, covering automatic speech recognition (ASR), Wu-to-Mandarin translation, speaker attribute prediction, speech emotion recognition, text-to-speech (TTS) synthesis, and instruction-following TTS (instruct TTS).
Furthermore, we release a suite of strong open-source models trained on WenetSpeech-Wu, establishing competitive performance across multiple tasks and empirically validating the effectiveness of the proposed dataset.
Together, these contributions lay the foundation for a comprehensive Wu dialect speech processing ecosystem, and we open-source proposed datasets, benchmarks, and models to support future research on dialectal speech intelligence~\footnote{\url{https://github.com/ASLP-lab/WenetSpeech-Wu-Repo}}.
\end{abstract}

\section{Introduction}
Speech processing has become a fundamental component of artificial intelligence, enabling natural and efficient human–machine interaction across a wide range of real-world applications~\citep{speech-under-survey, qwen3omni}.
For high-resource languages such as Mandarin Chinese and English, the speech processing ecosystem has reached a high level of maturity, characterized by large-scale and diverse datasets~\citep{wenetspeech, Heiga2019libritts, Haorui2024emilia}, publicly available and multi-dimensional evaluation benchmarks~\citep{mmau, mmsu, msubench}, and a growing number of powerful open-source models~\citep{funaudiollm, paraformer, whisper}.
Together, these components form a virtuous cycle in which data, benchmarks, and models mutually reinforce one another, continuously driving rapid progress in both academic research and practical deployment.
In contrast, the Wu dialect~\footnote{\url{https://en.wikipedia.org/wiki/Wu_Chinese}}, a crucial Chinese dialect that remains relatively underexplored in speech technology community, 
is associated with a severely underdeveloped speech processing ecosystem.
The lack of sufficient datasets, standardized benchmarks, and accessible models has substantially hindered both research advancement and real-world adoption.
To bridge this gap and promote inclusive speech technology, this work lays the foundation for a comprehensive speech processing ecosystem for the Wu dialect.
\begin{figure}[t]
  \centering
  \includegraphics[clip, trim=0.cm 1cm 13cm 0cm, width=1\linewidth]{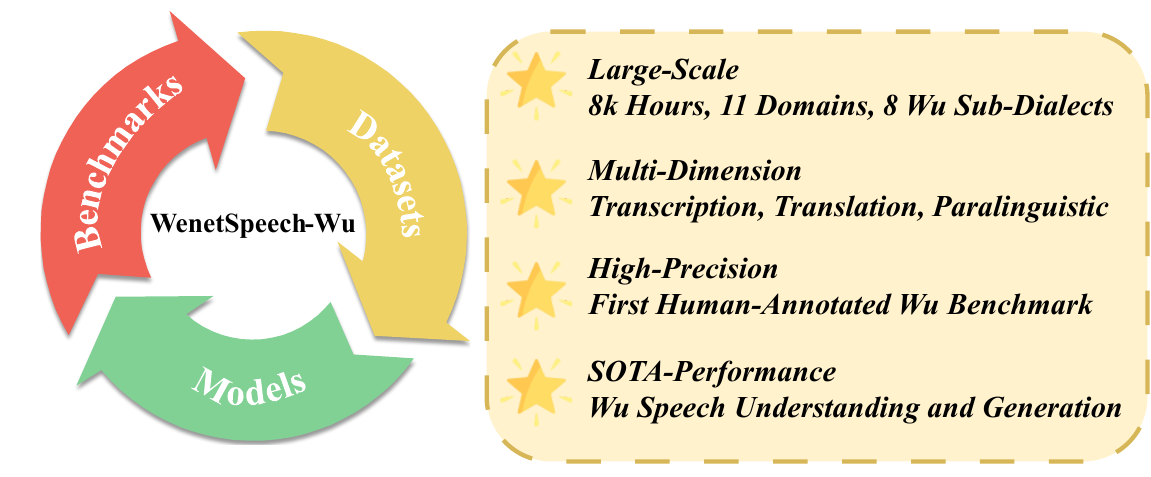}

  \caption{Highlights of WenetSpeech-Wu.}
  \label{fig:overall}
\end{figure}

\begin{table*}[thp]
\caption{
Comparison of typical low-resource speech processing resources related to WenetSpeech-Wu. 
$\cmark$, $\triangle$, and \protect\xmark~indicate experimentally verified availability, unverified availability, and absence of dimension, respectively.
Abbreviations: 
Data Comp (Data Composition; CN = China, SEA = Southeast Asia), 
Hrs (hours), 
Trans (transcription), 
Para (paralinguistics), 
Transl (translation), 
Q Tier (quality grading tier), 
Instr TTS (instruct TTS), 
Sp Und. (speech understanding), Wu-SH (Shanghai sub-dialect of Wu).
}
\setlength{\tabcolsep}{1.3mm}
\footnotesize

\centering
\begin{tabular}{
>{\raggedright\arraybackslash}m{2.7cm}
>{\raggedright\arraybackslash}m{1.8cm}
>{\centering\arraybackslash}p{0.5cm} 
>{\centering\arraybackslash}p{0.6cm} 
>{\centering\arraybackslash}p{0.4cm} 
>{\centering\arraybackslash}p{0.6cm} 
>{\centering\arraybackslash}p{0.6cm}
>{\centering\arraybackslash}p{0.4cm} 
>{\centering\arraybackslash}p{0.4cm} 
>{\centering\arraybackslash}p{0.4cm} 
>{\centering\arraybackslash}p{0.4cm} 
>{\centering\arraybackslash}p{0.7cm}
>{\centering\arraybackslash}p{0.4cm} 
>{\centering\arraybackslash}p{0.4cm} 
>{\centering\arraybackslash}p{0.7cm} 
>{\centering\arraybackslash}p{0.7cm}}
\toprule
& \multicolumn{6}{c}{\textbf{Dataset}} 
& \multicolumn{5}{c}{\textbf{Benchmark}} 
& \multicolumn{4}{c}{\textbf{Model}} \\
\cmidrule(lr){2-7} \cmidrule(lr){8-12} \cmidrule(lr){13-16}

\textbf{Resource} 
& Data Comp
& Hrs
& Trans
& Para
& Transl
& \makecell{Q\\Tier}

& ASR 
& AST
& Para 
& TTS
& \makecell{Instr\\TTS}

& ASR 
& TTS 
& \makecell{Sp\\Und}
& \makecell{Instr\\TTS} \\
\midrule

KeSpeech           
& 8 CN Accents & 1.5k &\cmark & $\triangle$ & \xmark & \xmark
& \cmark & \xmark & \xmark & \xmark & \xmark
& \cmark & \xmark & \xmark & \xmark \\

GigaSpeech2        
& 3 SEA Langs &30k & \cmark & \xmark & \xmark & \xmark & \cmark
& \xmark & \xmark & \xmark & \xmark
& \cmark & \xmark & \xmark & \xmark \\

WenetSpeech-Yue    
& CN Yue & 21k & \cmark & $\triangle$ & \xmark & \xmark
& \cmark & \xmark & \xmark & \cmark
& \xmark & \cmark & \cmark & \xmark & \xmark\\

WenetSpeech-Chuan  
& CN Sichuan &10k & \cmark & $\triangle$ & \xmark & \xmark
& \cmark & \xmark & \xmark & \cmark
& \xmark & \cmark & \cmark & \xmark & \xmark \\

Magicdata-Shanghai 
&  CN Wu-SH&4 & \cmark & \xmark & \xmark & \xmark
& \xmark & \xmark & \xmark & \xmark
& \xmark & \xmark & \xmark & \xmark & \xmark \\

\midrule
\textbf{WenetSpeech-Wu}     
&CN Wu & 8k& \cmark & \cmark & \cmark & \cmark
& \cmark & \cmark & \cmark & \cmark
& \cmark & \cmark & \cmark & \cmark & \cmark\\

\bottomrule
\end{tabular}

\label{tab:comparasion}
\end{table*}

The Wu dialect is a major branch of Chinese spoken by approximately 100 million speakers across Shanghai, Zhejiang, Jiangsu, and overseas communities, and is linguistically characterized by exceptional complexity.
It preserves the fully voiced consonant system inherited from Old Chinese. It features a highly intricate tone sandhi system in connected speech~\citep{wu-intro}, both of which pose substantial challenges for speech modeling.
Moreover, substantial variation exists among its sub-dialects, such as Shanghainese, Suzhounese, and Hangzhounese, further complicating the development of robust speech processing systems.

Despite its linguistic significance and broad speaker base, the Wu dialect remains severely under-resourced for speech processing.
From a data perspective, existing open-source resources are extremely limited in both scale and coverage: the only publicly available dataset, MagicData-Shanghai\footnote{\url{https://magichub.com/datasets/shanghai-dialect-conversational-speech-corpus/}}, provides merely 4.19 hours of annotated Shanghainese speech for automatic speech recognition (ASR), with no open datasets available for other Wu sub-dialects. 
Moreover, high-quality emotion annotations, speaker attributes and expressive speech–text pairs are entirely absent, which are essential for supporting a wide range of speech processing tasks.
From an evaluation perspective, the lack of publicly available benchmarks prevents fair comparison and systematic assessment across different methods. 
At the model level, even foundational speech processing systems, such as ASR and text-to-speech synthesis (TTS) models, are either unavailable or perform poorly, making both open-source and commercial systems essentially unusable for Wu dialect applications.
Collectively, these limitations underscore an urgent need to establish a comprehensive Wu dialect speech processing ecosystem, so as to facilitate dialectal speech research and foster the growth of the open-source community.

To address these challenges, we present a comprehensive open-source effort that lays the foundation for a Wu dialect speech processing ecosystem.
Specifically, we construct \textbf{WenetSpeech-Wu}, the first large-scale, multi-dimensionally annotated open-source corpus for the Wu dialect, comprising approximately 8,000 hours of eight Wu sub-dialects.
Building upon this dataset, we introduce \textbf{WenetSpeech-Wu-Bench}, the first standardized and publicly accessible benchmark designed for systematic evaluation of Wu dialect speech processing, covering ASR, Wu-to-Mandarin automatic speech translation (AST), speaker attribute prediction (gender and age), speech emotion recognition, TTS, and instruction-following TTS (instruct TTS).
Furthermore, we release a suite of strong open-source models trained on WenetSpeech-Wu, including ASR, TTS, unified speech understanding, and instruct TTS models, which substantially outperform existing open-source and commercial systems, thereby establishing strong models and empirically demonstrating the effectiveness of the proposed dataset.
Together, these contributions lay the foundation for a comprehensive, publicly accessible ecosystem for Wu dialect speech processing.

Our main contributions are threefold:
\begin{itemize}
    \item We release \textbf{WenetSpeech-Wu}, the first large-scale and multi-dimensionally annotated open-source corpus for the Wu dialect, containing approximately 8,000 hours of speech data.
    \item We introduce \textbf{WenetSpeech-Wu-Bench}, the first standardized benchmark for evaluating Wu dialect speech processing, covering ASR, Wu-to-Mandarin AST, speaker attribute prediction, speech emotion recognition,  TTS, and instruct TTS tasks.
    \item We open-source strong models trained on WenetSpeech-Wu, demonstrating substantial improvements over existing systems and validating the effectiveness of proposed dataset.
\end{itemize}

\section{Related Work}
\subsection{Speech Datasets for Low-Resource Languages and Chinese Dialects}
Recently, significant efforts have been devoted to constructing speech corpora for low-resource languages and dialects, aiming to mitigate data scarcity and promote inclusive speech technologies~\citep{thai}.
Representative large-scale corpora such as Common Voice~\citep{commonvoice}, GigaSpeech2~\citep{gigaspeech2}, and KeSpeech~\citep{kespeech} have substantially advanced research on speech processing for low-resource or underrepresented languages.
Within the Chinese dialect family, several dialect-oriented speech resources have been released recently, including WenetSpeech-Yue~\citep{ws-yue} for Cantonese and WenetSpeech-Chuan~\citep{ws-chuan} for Sichuanese.
However, these datasets primarily focus on ASR and, to a limited extent, TTS: Common Voice and GigaSpeech2 provide only transcriptions, KeSpeech augments transcriptions with dialect and basic speaker labels, and WenetSpeech-Yue and WenetSpeech-Chuan include paralinguistic annotations but evaluate their resources only through ASR and TTS experiments, without validating the effectiveness of the paralinguistic labels.
As a result, the usability of such resources for broader speech processing tasks, such as speech emotion recognition, AST, and instruct TTS, remains limited.

Despite its linguistic and practical importance, the Wu dialect remains severely under-resourced 
in terms of publicly available speech data.
To the best of our knowledge, MagicData-Shanghai is the only open-source Wu dialect dataset, providing merely 4.19 hours of annotated Shanghainese speech for ASR.
Such extreme data scarcity, together with the lack of diverse annotations has substantially hindered progress in Wu dialect speech processing research.

\subsection{Wu Dialect Speech Processing: Models and Evaluation}
Early studies on Wu dialect speech processing mainly focused on ASR or TTS for individual sub-dialects, most notably Shanghainese, while largely ignoring other major variants such as Suzhounese and Hangzhounese.
More recent works incorporate the Wu dialect as a minor component within large-scale multilingual ASR or TTS systems, including Dolphin~\citep{dolphin}, Qwen3-ASR, Qwen3-TTS, Step-Audio2~\citep{stepaudio2}, and Qwen3-Omni~\citep{qwen3omni}.
In these systems, Wu dialect speech is treated as a low-priority subset rather than a primary research target, resulting in foundational ASR and TTS models whose performance remains insufficient for practical Wu dialect applications.
Beyond ASR and TTS, speech understanding models and instructing TTS systems tailored to the Wu dialect remain largely unexplored.

More importantly, even for basic ASR and TTS tasks, there is currently no publicly available and standardized benchmark for evaluation.
As a consequence, existing studies typically evaluate models on private test sets, severely hindering fair comparison, reproducibility, and systematic progress.

In summary, as illustrated in Tabel~\ref{tab:comparasion}, these limitations motivate the introduction of WenetSpeech-Wu, which aims to establish a complete Wu dialect speech processing ecosystem by jointly advancing datasets, benchmarks, and models.

\section{WenetSpeech-Wu}
\subsection{Data Construction Pipeline}
\begin{figure*}[t]
  \centering
  \includegraphics[width=1.0\linewidth]{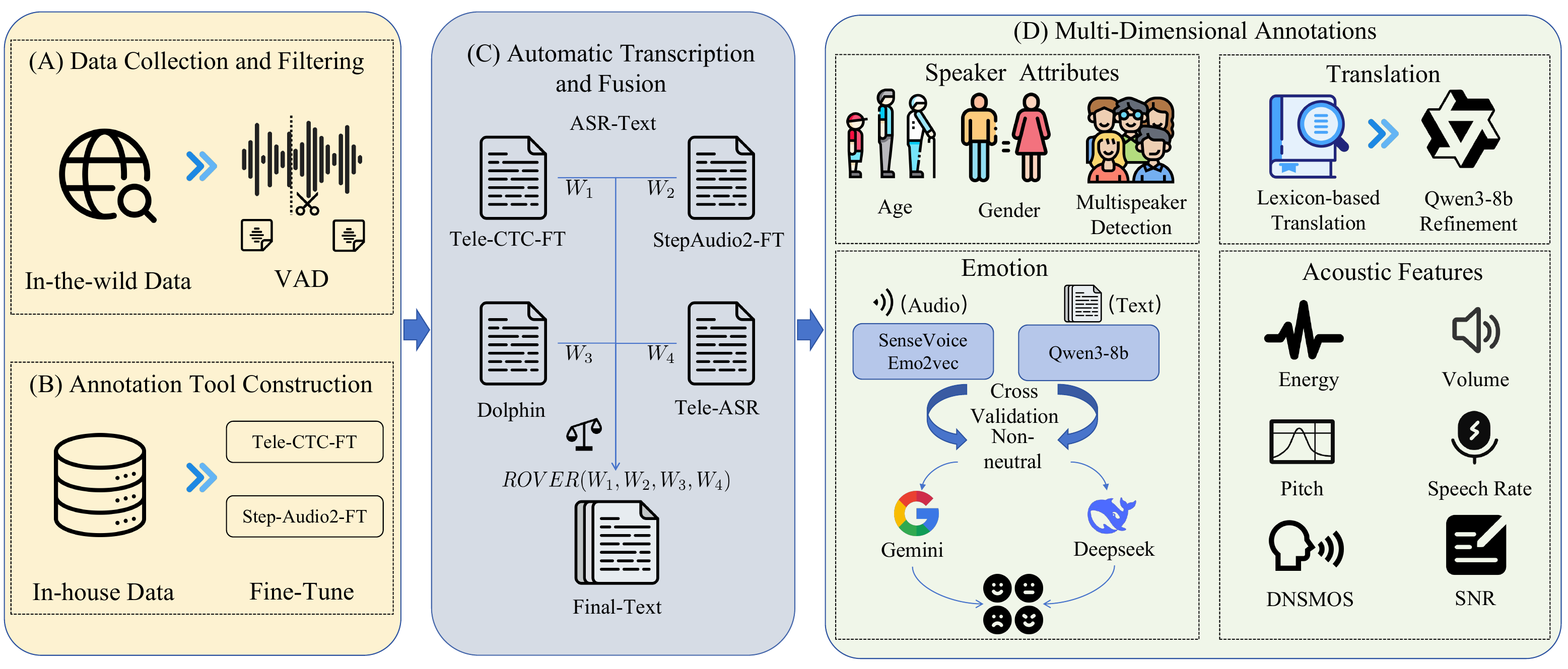}
  \caption{Data construction pipeline for WenetSpeech-Wu.}
  \label{fig:pipeline}
\end{figure*}

We propose an automatic and scalable pipeline for constructing a large-scale Wu dialect speech dataset with multi-dimensional annotations, as illustrated in Figure~\ref{fig:pipeline}. 
The pipeline is designed to enable efficient data collection, robust automatic transcription, and diverse downstream annotations.

\textbf{Data Collection and Filtering.} 
We collect large-scale in-the-wild Wu dialect speech from diverse domains and sub-dialects. Non-Wu data are first removed based on metadata filtering, followed by WebRTC VAD-based segmentation. We further apply quality filtering using DNSMOS and signal-to-noise ratio (SNR), resulting in a high-quality speech corpus.

\textbf{Annotation Tool Construction.}
To support large-scale automatic transcription, we fine-tune two pretrained ASR models using 880 hours of manually annotated Wu dialect speech.
Specifically, we fine-tune Tele-CTC-FT\footnote{\url{https://huggingface.co/Tele-AI/TeleSpeech-ASR1.0}}, a Connectionist Temporal Classification (CTC)-based dialect self-supervised learning (SSL) model, and Step-Audio2-FT~\citep{stepaudio2}.
These models are used as complementary automatic annotators for Wu dialect transcription.

\textbf{Automatic Transcription and Fusion.} 
We adopt Recognizer Output Voting Error Reduction (ROVER)~\citep{rover} to fuse transcription from multiple ASR systems.
Specifically, we combine the outputs from the two fine-tuned Wu dialect ASR models, Dolphin~\citep{dolphin} and TeleASR~\citep{telespeechpt}, and determine the model weights via grid search.
The fused results provide final transcriptions with confidence scores.

\textbf{Multi-Dimensional Annotations.} 
Each speech sample is annotated across multiple dimensions. Speaker attributes, including gender and age, are inferred using the VoxProfile~\citep{voxprofile}, while multi-speaker presence is detected using Pyannote~\citep{pyannote}. Wu-to-Mandarin translations are generated through lexicon-based mapping and further refined using Qwen3-8B~\citep{qwen3}, a large language model, aiming to provide fluent standard Mandarin references. 
Emotion annotations are obtained through a multi-stage, cross-modal procedure.
Initial predictions are produced using SenseVoice~\citep{funaudiollm} and Emo2Vec~\citep{emo2vec} for acoustic signals, and Qwen3-8B for textual content. Samples jointly identified as non-neutral are further analyzed using DeepSeek-R1 based on text and Gemini-2.5-Pro based on acoustic information, with the final label determined by the intersection of the two.
In addition, prosodic acoustic features, including speech rate, loudness, energy, and pitch, are extracted by Dataspeech~\citep{deepspeeh} to support speech generation tasks.

\begin{figure*}[!htb]
\centering
\includegraphics[width=\textwidth]{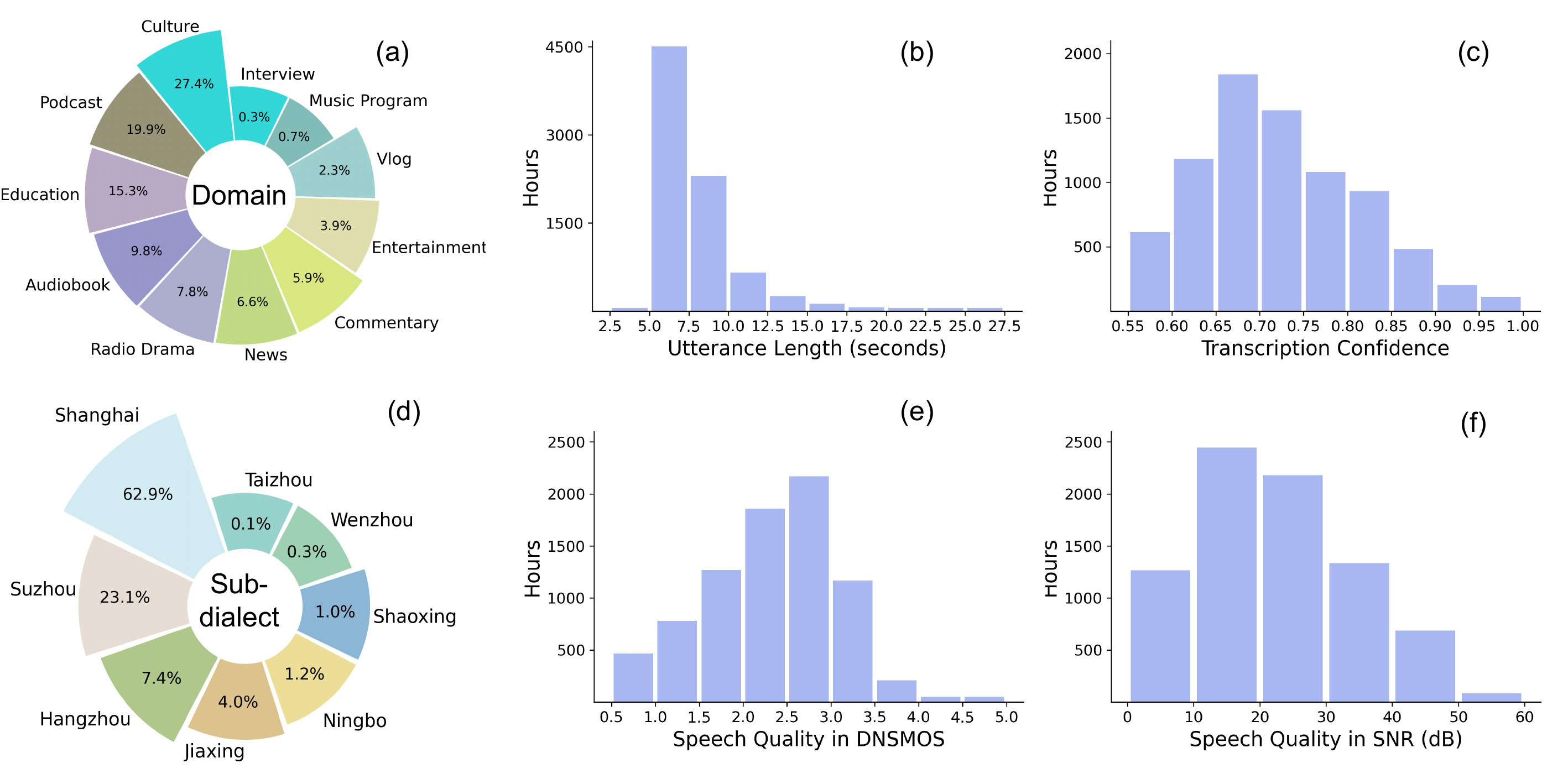}
\caption{Statistical overview of WenetSpeech-Wu.}
\label{fig:overviewofdataset}
\end{figure*}

\subsection{Datasets}

WenetSpeech-Wu is the first large-scale Wu dialect speech corpus with multidimensional annotations.
It contains rich metadata and annotations, including transcriptions with confidence scores, Wu-to-Mandarin translations, domain and sub-dialect labels, speaker attributes, emotion annotations, and audio quality measures.
The dataset comprises approximately 8,000 hours of speech collected from diverse domains and covers eight Wu sub-dialects.
To support a wide range of speech processing tasks with heterogeneous quality requirements, we further adopt a task-specific data quality grading strategy.
In the following, we describe the dataset statistics, domain and dialect coverage, annotation distributions, and quality control mechanisms in detail.

\textbf{Duration and Confidence Distribution.} 
WenetSpeech-Wu contains 8,000 hours of speech with 3.86M utterances, with utterance durations up to 30 seconds and an average duration of 7.45 seconds.
We use the transcription confidence produced by the weighted ROVER as a measure of annotation quality, and retain utterances with confidence scores above 0.55.
The detailed distributions of utterance duration and transcription confidence are shown in Figure~\ref{fig:overviewofdataset}b and Figure~\ref{fig:overviewofdataset}c.

\textbf{Domain and Sub-Dialect Coverage.}
WenetSpeech-Wu covers a wide range of speech domains and Wu sub-dialects.
The domains include \emph{News, Culture, Vlog, Entertainment, Education, Podcast, Commentary, Interview, Radio Drama, Music Program, and Audiobook}, with their distribution shown in Figure~\ref{fig:overviewofdataset}a.
In terms of dialectal coverage, approximately 37\% of the recordings cannot be reliably assigned to a specific Wu sub-dialect and are therefore labeled as Unknown. The remaining recordings span multiple identified Wu sub-dialects, including \emph{Shanghainese, Suzhounese, Shaoxingnese, Ningbonese, Hangzhounese, Jiaxingnese, Taizhounese, and Wenzhounese}, whose distribution is illustrated in Figure~\ref{fig:overviewofdataset}d.

\textbf{Audio Quality.}
As shown in Figure~\ref{fig:overviewofdataset}e and Figure~\ref{fig:overviewofdataset}f, most utterances have SNRs between 10 and 40 dB, with a peak at 20–30 dB.
MOS scores are primarily in the range 2.0–3.5.






\begin{table}[t]
\centering
\caption{Statistics of speaker attributes and emotion annotations in WenetSpeech-Wu.}
\label{tab:attr_emotion_stats}

\footnotesize
\setlength{\tabcolsep}{1.5mm}

\begin{tabular}{
>{\centering\arraybackslash}p{1.2cm}
>{\centering\arraybackslash}p{1.8cm}
>{\centering\arraybackslash}p{1.2cm}
}
\toprule
\textbf{Task} & \textbf{Category} & \textbf{Hours} \\
\midrule

\multirow{2}{*}{Gender}
 & Male   & 4,135 \\
 & Female & 1,331 \\
\midrule

\multirow{4}{*}{Age}
 & Teenagers     & 372 \\
 & Youth         & 1,673 \\
 & Middle-aged   & 2,003 \\
 & Elderly       & 1,418 \\
\midrule

\multirow{5}{*}{Emotion}
 & Neutral   & 5,102 \\
 & Happy     & 73 \\
 & Sad       & 81 \\
 & Angry     & 109 \\
 & Surprised & 101 \\
\bottomrule
\end{tabular}
\label{emo_bench}
\end{table}

\textbf{Speaker Attributes and Emotion Annotations.}
We annotate gender, age, and emotion labels for single-speaker segments.
Gender is categorized into \emph{Male} and \emph{Female}, age into four groups including 0–17 as \emph{Teenagers}, 18–35 as \emph{Youth}, 36–59 as \emph{Middle-aged}, and 60+ as \emph{Elderly}, and emotion into five classes: \emph{Neutral, Happy, Sad, Surprised, and Angry}.
The distribution of each category is shown in Table~\ref{emo_bench}.

\textbf{Task-Specific Data Quality Grading.}
To support practical training across heterogeneous speech tasks, we adopt a task-specific data-quality grading strategy aligned with the task-specific quality requirements.
For ASR and TTS, we construct two quality tiers.
The normal-quality subset is designed for large-scale pretraining and 
prioritizes data coverage and diversity, requiring only moderate transcription confidence.
In contrast, the high-quality subset targets supervised fine-tuning (SFT) 
and applies stricter filtering criteria, including high transcription confidence, 
clean acoustic conditions, and reliable speaker segmentation, 
to provide strong and stable supervision.
For tasks that are more sensitive to annotation noise and semantic ambiguity, 
including Wu-to-Mandarin AST, speaker attribute prediction, speech emotion recognition, TTS, and instruct TTS, we adopt stringent selection criteria, such as single-speaker recordings, high MOS, SNR, and pitch standard deviation, and verified annotation consistency, as shown in Table~\ref{tab:quality}.

\begin{table}[!t] 
\centering

\caption{Task-specific data selection and quality tiers in WenetSpeech-Wu. Abbreviations: Q Tier (quality grading tier), Text Conf (transcription confidence), Spk (speaker type, Mono = single-speaker), Pitch Std. (pitch standard deviation), Expr (expressive data obtained via emotion label), Spk Attr (speaker attributes), Emo (emotion), Inst Pro (instruct TTS prosodic control), Inst Emo (instruct TTS emotional control).}


\label{tab:quality}
\footnotesize 
\setlength{\tabcolsep}{1.5mm}

\begin{tabular}{
>{\centering\arraybackslash}p{0.6cm}
>{\centering\arraybackslash}p{0.6cm}
>{\centering\arraybackslash}p{0.6cm}
>{\centering\arraybackslash}p{1cm}
>{\centering\arraybackslash}p{0.6cm}
>{\centering\arraybackslash}p{0.5cm}
>{\centering\arraybackslash}p{0.79cm}
>{\centering\arraybackslash}p{0.7cm}
}

\toprule
\textbf{Task} &
\textbf{Hrs} &
\textbf{\makecell{Q\\Tier}} &
\makecell[c]{\textbf{Text} \\ \textbf{Conf}} &
\makecell[c]{\textbf{Spk}} &
\textbf{SNR} &
\makecell[c]{\textbf{Pitch} \\ \textbf{Std.}} &
\makecell[c]{\textbf{Expr}} \\
\midrule

\multirow{2}{*}{ASR}
 & 7388 & Mid & > 0.6  & -- & -- & -- & -- \\
 & 795  & High   & > 0.85 & -- & -- & -- & -- \\
\midrule

AST & 795 & High & $>0.85$ & -- & -- & -- & -- \\
\midrule

\makecell{Spk\\Attr} & 2986 & High & -- & Mono & -- & -- & -- \\
\midrule

Emo & 500 & High & -- & Mono & >10 & >50 & \checkmark \\

\midrule
\multirow{2}{*}{TTS}
 & 7388 & Mid & > 0.6 & -- & -- & -- & -- \\
 & 1500 & High & > 0.65 & Mono & >10 & >50 & -- \\

\midrule

\makecell{Inst\\Pro}
& 679 & High & > 0.7 & Mono & >30 & -- & -- \\
\midrule
\makecell{Inst\\Emo} & 161 & High & > 0.7 & Mono & >10 & >50 & \checkmark \\

\bottomrule
\end{tabular}
\end{table}

\section{WenetSpeech-Wu-Bench}
We introduce \textbf{WenetSpeech-Wu-Bench}, the first publicly available, manually curated benchmark for Wu dialect speech processing, covering ASR, Wu-to-Mandarin AST, speaker attributes, emotion recognition, TTS, and instruct TTS, and providing a unified platform for fair evaluation.

\textbf{Automatic Speech Recognition.}
The ASR test set of WenetSpeech-Wu-Bench comprises 9.75 hours of Shanghainese, Suzhounese, and Mandarin code-mixed speech, including single-speaker and multi-speaker scenarios. Performance is evaluated using character error rate (CER), with character-level errors reported for detailed analysis.

\textbf{Wu-to-Mandarin Speech Translation.}
We construct a Wu-to-Mandarin AST test set.
The test set contains 3000 Wu dialect utterances totaling 4.4 hours with manually verified standard Mandarin translations, covering multiple domains.
Translation quality is evaluated using the BLEU score~\footnote{\url{https://github.com/mjpost/sacrebleu}}.

\textbf{Speaker Attribute Prediction and Speech Emotion Recognition.}
This test set evaluates the prediction of age, gender, and emotion for Wu dialect speech.
For speaker attributes, gender is coded as male or female, with 1,500 samples per class. 
And age into four groups: teenagers aged 17 and under, youth aged 18 to 35, middle-aged adults aged 36 to 59, and elderly aged 60 and above, with 500 samples in each group.
For emotion, there are 300 neutral samples, 200 happy samples, 100 sad samples, 200 angry samples, and 200 surprised samples, for a total of 1,000 samples. 
Performance is measured by category-wise and overall classification accuracy.

\textbf{Text to Speech.}
We constructed a TTS testset as part of WenetSpeech-Wu-Bench, comprising 144 easy and 98 hard test sentences.
The texts were further reviewed and refined by professional Wu dialect experts. Prompt audio samples were selected from the open-source Magicdata-Shanghai, and 12 speakers of the Wu dialect were selected using strict filtering criteria.
For evaluation, speaker similarity is measured using WeSpeaker \footnote{\url{https://github.com/wenet-e2e/wespeaker}} based speaker embedding similarity, while intelligibility is assessed by computing CER with our proposed Step-Audio2-Wu-ASR model. Additionally, subjective listening tests are conducted, including intelligibility MOS (IMOS), similarity MOS (SMOS), and accent MOS (AMOS). 
The subjective evaluation was conducted with 23 listeners, each rating 20 selected samples.

\textbf{Instruct TTS.} 
WenetSpeech-Wu-Bench includes two evaluation sets for instruct TTS. 
The prosodic control test set consists of five speech prompts, each spoken at a moderate speaking rate and normal fundamental frequency, and synthesized into 20 sentences with controlled variations in speaking rate and pitch. 
For evaluation, two experiments are conducted: one with a fast speaking rate and high pitch, and another with a slow rate and low pitch. 
These samples are automatically annotated using Dataspeech~\citep{deepspeeh}. 
Each pair is scored as one if the relative speaking rate and pitch match the intended instructions, and zero otherwise. 
The prosodic classification metric is the average score across all pairs, reflecting the model's ability to follow prosodic instructions~\citep{10389672}.
The emotional control test set evaluates a model’s ability to follow emotion-related instructions. 
We select 10 reference prompts that contain no explicit emotional expression. 
Based on each prompt, 50 sentences are synthesized for each of the four target emotions: anger, sadness, happiness, and surprise. 
The samples are evaluated using our Step-Audio2-Wu-Und model. 
A sample is considered correct if the predicted emotion matches the intended target emotion, and the mean classification accuracy is reported as the evaluation metric~\cite{gao2025differentiablerewardoptimizationllm}.
Additionally, subjective listening tests are conducted to assess the quality of instruction-following. 
Listeners rate the samples using two scores: the prosodic MOS (PMOS)~\cite{chan2025objectiveinterpretableprosodyevaluation} and emotional Mean Opinion Score (EMOS)~\cite{cho2025emosphereemotioncontrollablezeroshottexttospeech}. The evaluation involves 23 listeners, each rating 15 samples, to provide perceptual judgments on how well the synthesized speech follows the intended emotional instructions.

\section{Models \& Experiments}
To address the lack of Wu dialect speech processing models, we develop models for the two core aspects of speech processing, speech understanding and speech generation trained on WenetSpeech-Wu. 
These include ASR models and unified speech understanding models for understaning, as well as TTS and instruct TTS models for generation.
\subsection{Speech Understanding}

\begin{table}[!t]
\caption{ASR results (CER\%) on various test sets. 
\colorbox{gray!8}{Gray}, \colorbox{red!8}{red}, \colorbox{green!10}{light green}, and \colorbox{green!20}{dark green} rows denote open-source baselines, commercial models, ASR models trained on WenetSpeech-Wu, and annotation models trained on in-house data. 
\textbf{Bold} numbers indicate best results; \underline{underlined} numbers indicate second-best results.}
\setlength{\tabcolsep}{1mm}
\centering
\footnotesize

\begin{tabular}{l*{4}{c}}
\toprule

\multirow{2}{*}{Model} 
& \multicolumn{2}{c}{In-House} 
& \multicolumn{1}{c}{WS-Wu-Bench} \\
\cmidrule(lr){2-3} \cmidrule(lr){4-4}
& Dialogue & Reading & ASR \\
\midrule

\textbf{ASR Models} & & & \\
\rowcolor{gray!10} Paraformer & 63.13 & 66.85 & 64.92 \\
\rowcolor{gray!10} SenseVoice-small & 29.20 & 31.00 & 46.85 \\
\rowcolor{gray!10} Whisper-medium & 79.31 & 83.94 & 78.24 \\
\rowcolor{gray!10} FireRedASR-AED-L & 51.34 & 59.92 & 56.69 \\
\rowcolor{gray!10} Step-Audio2-mini & 24.27 & 24.01 & 26.72 \\

\rowcolor{red!10} Qwen3-ASR & 23.96 & 24.13 & 29.31 \\
\rowcolor{red!10} Tencent-Cloud-ASR & 23.25 & 25.26 & 29.48 \\
\rowcolor{red!10} Gemini-2.5-pro & 85.50 & 84.67 & 89.99 \\

\rowcolor{green!10} Conformer-U2pp-Wu & 15.20 & 12.24 & 15.14 \\
\rowcolor{green!10} Whisper-medium-Wu & 14.19 & 11.09 & \underline{14.33} \\
\rowcolor{green!10} Step-Audio2-Wu-ASR & \underline{8.68} & 7.86 & \textbf{12.85} \\

\hline
\textbf{Annotation Models} & & & \\
\rowcolor{gray!10} Dolphin-small & 24.78 & 27.29 & 26.93 \\
\rowcolor{red!10} TeleASR & 29.07 & 21.18 & 30.81 \\
\rowcolor{green!20} Step-Audio2-FT & \textbf{8.02} & \textbf{6.14} & 15.64 \\
\rowcolor{green!20} Tele-CTC-FT & 11.90 & \underline{7.23} & 23.85 \\

\bottomrule
\end{tabular}

\label{tab:asr_results}
\end{table}
\textbf{ASR models.}
To accommodate different application scenarios, we develop three Wu dialect ASR models at three scales: a Conformer-U2pp-Wu~\citep{u2pp} model, a Whisper-Medium-Wu~\citep{whisper} model, and a Step-Audio2-Wu-ASR model. 
The Conformer-U2pp-Wu and Whisper-Medium-Wu models are trained within the Wenet framework~\cite{wenet}, with Conformer-U2pp-Wu trained from scratch, and Whisper-Medium-Wu fine-tuned from publicly available pre-trained weights. 
The Step-Audio2-Wu-ASR model is a Step-Audio2-mini fine-tuned in a parameter-efficient manner using LoRA~\citep{lora} within the MS-Swift framework~\citep{swift}.
All models are pretrained on the ASR-Mid subset and conduct SFT on the ASR-High subset.
Evaluation is performed on the ASR test set of WenetSpeech-Wu-Bench as well as two in-house manually annotated test sets covering dialogue and reading scenarios, enabling comprehensive assessment across diverse speaking conditions.
\begin{table}[!t]
\caption{Speech understanding performance on WenetSpeech-Wu-Bench.}
\setlength{\tabcolsep}{0.7mm}
\centering
\footnotesize

\begin{tabular}{l*{5}{c}}
\toprule
Model & ASR & AST & Gender & Age & Emotion \\
\midrule
\rowcolor{gray!10} Qwen3-Omni
& 44.27 & 33.31 & \textbf{0.977} & \underline{0.541} & \underline{0.667} \\

\rowcolor{gray!10} Step-Audio2-mini
& \underline{26.72} & \underline{37.81} & 0.855 & 0.370 & 0.460 \\

\rowcolor{green!10} Step-Audio2-Wu-Und
& \textbf{13.23} & \textbf{53.13} & \underline{0.956} & \textbf{0.729} & \textbf{0.712} \\
\bottomrule
\end{tabular}

\label{tab:speech_understanding}
\end{table}

\begin{table*}[t]
\footnotesize
\centering
\caption{
TTS results on WenetSpeech-Wu-Bench. \textbf{Bold} and \underline{underlined} values denote the best and second-best results, respectively; \colorbox{green!8}{light green} rows indicate models trained on WenetSpeech-Wu or further fine-tuned on an internal high-quality dataset.}
\setlength{\tabcolsep}{1pt}
\renewcommand{\arraystretch}{1.1}
\begin{tabular}{lcccccccccc}
\toprule
\multirow{2}{*}{\textbf{Model}} & \multicolumn{5}{c}{\textbf{WS-Wu-Eval-TTS-easy}} & \multicolumn{5}{c}{\textbf{WS-Wu-Eval-TTS-hard}} \\ 
\cmidrule(lr){2-6} \cmidrule(lr){7-11}
& \textbf{CER(\%)$\downarrow$} & \textbf{SIM$\uparrow$} & \textbf{IMOS$\uparrow$} & \textbf{SMOS$\uparrow$} & \textbf{AMOS$\uparrow$} & \textbf{CER(\%)$\downarrow$} & \textbf{SIM$\uparrow$} & \textbf{IMOS$\uparrow$} & \textbf{SMOS$\uparrow$} & \textbf{AMOS$\uparrow$} \\ 
\midrule
Qwen3-TTS{$\dagger$} & \underline{5.95} & -- & \underline{4.35} & -- & \underline{4.19}  & \underline{16.45} & -- & \underline{4.03} & -- & \textbf{3.91} \\
DiaMoE-TTS    & 57.05 & 0.702 & 3.11 & 3.43 & 3.52 & 82.52 & 0.587 & 2.83 & 3.14 & 3.22 \\
CosyVoice2    & 10.33 & 0.713 & 3.83 & 3.71 & 3.84 & 82.49 & \underline{0.618} & 3.24 & 3.42 & 3.37 \\
\rowcolor{green!8}
CosyVoice2-Wu-CPT  & 6.35 & \textbf{0.727} & 4.01 & \textbf{3.84} & 3.92
& 32.97 & \textbf{0.620} & 3.72 & \textbf{3.55} & 3.63 \\
\rowcolor{green!8}
CosyVoice2-Wu-SFT   & 6.19 & \underline{0.726} & 4.32 & \underline{3.78} & 4.11
& 25.00 & 0.601 & 3.96 & \underline{3.48} & 3.76 \\
\rowcolor{green!8}
CosyVoice2-Wu-SS{*} & \textbf{5.42} & -- & \textbf{4.37} & -- & \textbf{4.21}
 & \textbf{15.45} & -- & \textbf{4.04} & -- & \underline{3.88} \\
\bottomrule
\end{tabular}

\begin{tablenotes}
\footnotesize
\item{$\dagger$} Commercial system with a single fixed speaker, and speaker similarity is not considered.\item{*} Single-speaker finetuned model, and speaker similarity is not evaluated.

\end{tablenotes}
\label{tab:comparison}
\label{tab:wu_tts_results}
\end{table*}

For comparison, we include both open-source and commercial baselines, such as Dolphin, SenseVoice, Paraformer~\citep{paraformer}, FireRedASR~\citep{fireredasr} as an open-source baseline, Qwen3-ASR, Gemini-2.5-Pro, Tele-ASR~\citep{telespeechpt}, and Tencent-Cloud-ASR as a commercial baseline. 
The two in-house test sets are strictly out-of-set evaluations for Conformer-U2pp-Wu, Whisper-medium-Wu, and Step-Audio2-Wu-ASR.
Conversely, the WS-Wu-Bench-ASR test set serves as an out-of-set evaluation for Step-Audio2-FT and Tele-CTC-FT, which are trained exclusively on in-house data.

As shown in Table~\ref{tab:asr_results}, all existing open-source and commercial ASR systems perform poorly across all three test sets, indicating that they are not viable for Wu dialect recognition. 
In contrast, models trained on WenetSpeech-Wu achieve state-of-the-art performance across all model scales, with even the smallest Conformer-U2pp-Wu substantially outperforming all prior systems.

\textbf{Speech understanding models.}
For speech understanding, we train the Step-Audio2-Wu-Und model on the WenetSpeech-Wu corpus using task-aware, quality-graded data. 
The model is first pretrained on ASR-Mid and High-quality AST and age, gender, and emotion annotations, and then fine-tuned with the ASR-high subset alongside the same non-ASR tasks.

For benchmarking, we compare against baseline models, including Step-Audio2-mini and Qwen3-Omni, for speech understanding tasks. 
As shown in Table~\ref{tab:speech_understanding}, after multi-task fine-tuning, the ASR performance of Step-Audio2-Wu-Und shows a slight drop than Step-Audio2-Wu-ASR but still achieves the second-best results. For the Wu-to-Mandarin AST task, it significantly outperforms baseline models. Comparison with Step-Audio2-mini illustrates the mismatch between Wu dialect and Mandarin in gender, age, and emotion prediction, which is effectively addressed by our data. Compared with Qwen3-Omni, our model shows notable improvements in age and emotion prediction, while performing slightly lower on gender classification.


\subsection{Speech Generation}
\textbf{TTS models.}
In the continual pre-training (CPT) stage, we continue training from the open-source CosyVoice2~\citep{cosy2} checkpoint using the TTS-Mid dataset for ten epochs. 
In the SFT stage, we use TTS-High dataset trained for three epochs. Finally, we perform speaker-specific fine-tuning on a 10-hour internal high-quality dataset.


As shown in Table~\ref{tab:wu_tts_results}, the experimental results demonstrate that the staged training strategy significantly improves CosyVoice2’s speech synthesis performance in terms of Wu dialect capability. The CPT stage, leveraging large-scale pipeline-processed data, enhances the model’s fundamental capabilities and robustness, particularly on challenging samples. The SFT stage, further improves speech naturalness and expressiveness. Finally, the Single-Speaker Supervised Fine-Tuning (SS-SFT) stage achieves the best performance across CER, IMOS, and AMOS metrics. Overall, CosyVoice2-Wu-SS approaches or surpasses the baseline Qwen3-TTS, DiaMoE-TTS \cite{chen2025diamoettsunifiedipabaseddialect}, and CosyVoice2 in most metrics, especially in difficult sample synthesis.

\textbf{Instruct TTS models.}
The instruction training data are from the Inst Pro and Inst Emo datasets as introduced in Table~\ref{tab:quality}. Training was performed on the instruction text and tags using a small learning rate for five epochs. Compared with the model before instruction fine-tuning, the results showed clear improvements across all controllability metrics on WenetSpeech-Wu-Bench, as shown in Table~\ref {tab:instruct_emotion}, and subjective listening tests confirmed strong perceptual control effects, validating the effectiveness of proposed data.
\begin{table}[!t]
\caption{Performance of instruct TTS model.}
\setlength{\tabcolsep}{0.8mm}
\centering
\footnotesize

\begin{tabular}{l c c c}
\toprule
\textbf{Type} & \textbf{Metric} & \textbf{\makecell{CosyVoice2-Wu\\-SFT}} & \textbf{\makecell{CosyVoice2-Wu\\-instruct}} \\
\midrule
\multirow{4}{*}{Emotion} 
 & Happy$\uparrow$ & 0.87 & \textbf{0.94} \\
 & Angry$\uparrow$ & 0.83 & \textbf{0.87} \\
 & Sad$\uparrow$ & 0.84 & \textbf{0.88} \\
 & Surprised$\uparrow$ & 0.67 & \textbf{0.73} \\
  \cmidrule(lr){2-4}
 & EMOS$\uparrow$ & 3.66 & \textbf{3.83} \\
\midrule
\multirow{3}{*}{Prosody} 
 & Pitch$\uparrow$ & 0.24 & \textbf{0.74} \\
 & Speech Rate$\uparrow$ & 0.26 & \textbf{0.82} \\
  \cmidrule(lr){2-4}
 & PMOS$\uparrow$ & 2.13 & \textbf{3.68} \\
\bottomrule
\end{tabular}
\label{tab:instruct_emotion}
\end{table}

\section{Conclusion}
In this work, we establish a Wu dialect speech processing ecosystem encompassing datasets, benchmarks, and models. 
We introduce WenetSpeech-Wu, an 8,000-hour large-scale corpus with rich multi-dimensional annotations across multiple Wu sub-dialects, and present the first public Wu dialect benchmark covering ASR, Wu-to-Mandarin AST, speech attribute and emotion recognition, TTS, and instruct TTS. 
Building on this dataset, we release a suite of ASR, TTS, unified speech understanding models, and instruct TTS models, enabling community research and demonstrating the effectiveness of WenetSpeech-Wu.


\section*{Limitations}

Despite the contributions of WenetSpeech-Wu to the dialectal speech processing community, several limitations remain to be addressed in future work.
Although WenetSpeech-Wu is large-scale and covers multiple Wu sub-dialects, the data distribution across dialects and domains is not perfectly balanced, which may affect model generalization to less-represented varieties.
In addition, many annotations are produced through automated or semi-automated pipelines. While we apply stringent quality control and filtering strategies, these annotations may still contain noise compared to fully manual labeling.
Finally, our baseline models are intended to provide strong and reproducible reference points rather than task-optimal solutions, and future work may further improve performance through more specialized modeling and training strategies.

\bibliography{custom}

\appendix

\section{Appendix}
\subsection{Details of Datasets}

\begin{figure}[hp]
\centering
\includegraphics[width=0.65\linewidth]{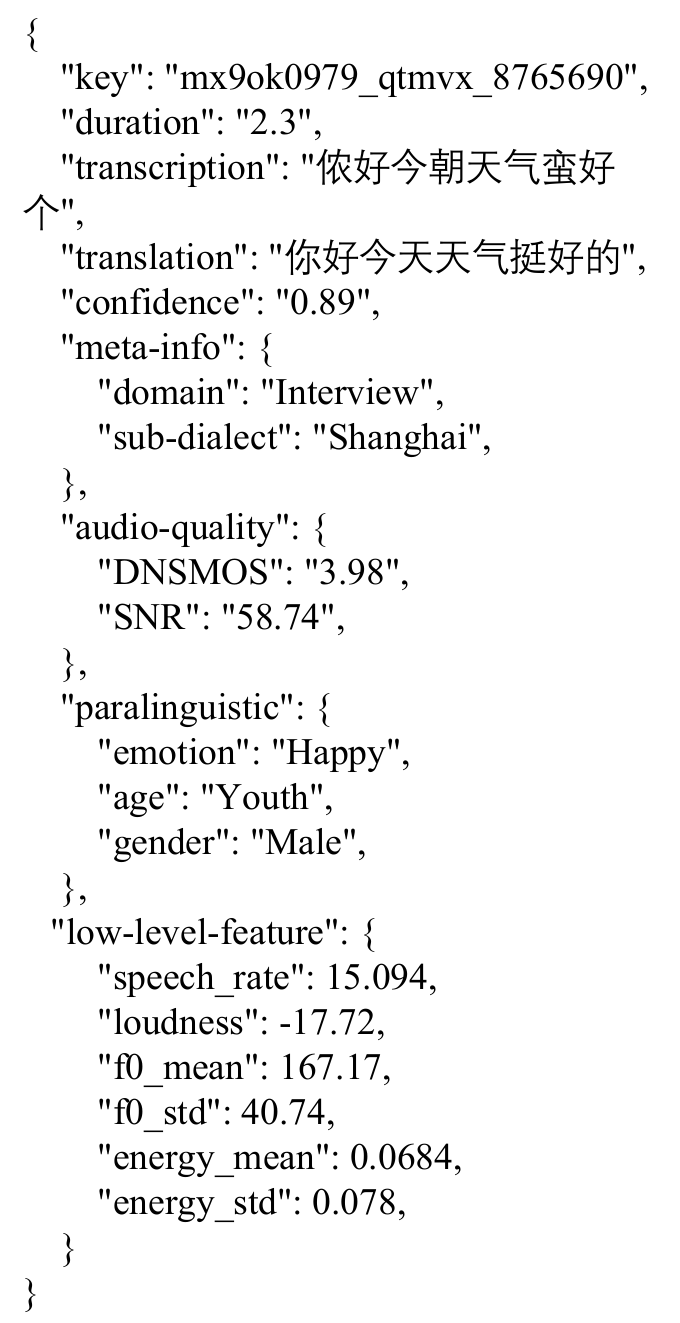}
\caption{Example of an Annotated Data Entry in JSON format}
\label{fig:dataset_details}
\end{figure}

\textbf{Example of an Annotated Data.}
Each speech sample is represented using a unified JSON format.
An entry includes a unique key, audio duration, Wu dialect transcription with its Mandarin translation, and an associated confidence score.
Contextual metadata, such as domain and sub-dialect, audio quality indicators (e.g., DNSMOS and SNR), and paralinguistic annotations (emotion, age, and gender) are organized into structured fields.
In addition, prosodic acoustic features, including speech rate, loudness, fundamental frequency statistics, and energy statistics, are provided to support fine-grained acoustic analysis.
Figure~\ref{fig:dataset_details} illustrates an example of the annotated JSON entry.

\begin{table}[bhp]
\caption{Statistics of the WenetSpeech-Wu-Bench.}
\footnotesize
\centering
\setlength{\tabcolsep}{1.5mm}

\begin{tabular}{l l c c}
\hline
\textbf{Task} & \textbf{Category} & \textbf{Utterances} & \textbf{Duration (h)} \\

\hline
ASR & - & 4851 & 9.75 \\
AST & - &   3000 & 4.4 \\
\hline

\multirow{2}{*}{Gender} 
 & Male   & 1500 & 2.10 \\
 & Female &   1500 & 2.29 \\
\cdashline{2-4}
 & Total  & 3000 & 4.39 \\
\hline

\multirow{4}{*}{Age} 
 & Youth              &   500 & 0.69 \\
 & Middle-aged & 500 & 0.79 \\
 & Elderly            &   500 & 0.69 \\
\cdashline{2-4}
 & Total & 1500 & 2.22 \\
\hline

\multirow{5}{*}{Emotion} 
 & Neutral   & 300 & 0.50 \\
 & Happy     &  200 & 0.23 \\
 & Sad       &  100 & 0.26 \\
 & Angry     &  200 & 0.15 \\
 & Surprised &  200 & 0.27 \\
\cdashline{2-4}
 & Total & 1000 & 1.41 \\
\hline

\end{tabular}

\label{tab:bench_stats}
\end{table}

\subsection{Statistics of the WenetSpeech-Wu-Bench} 
As summarized in Table~\ref{tab:bench_stats}, the WenetSpeech-Wu Bench consists of diverse test sets tailored for multiple speech-related tasks. For each task, we report the number of utterances and the corresponding total duration in hours.
\begin{table*}[t]
\centering
\setlength{\tabcolsep}{2mm}
\footnotesize
\caption{Optimization and training hyperparameters for proposed ASR and TTS models.}
\label{tab:training_hyperparams}
\begin{tabular}{
>{\raggedright\arraybackslash}m{4cm}
>{\centering\arraybackslash}m{1.2cm}
>{\centering\arraybackslash}m{1.6cm}
>{\centering\arraybackslash}m{1.6cm}
>{\centering\arraybackslash}m{1.6cm}
>{\centering\arraybackslash}m{0.6cm}
>{\centering\arraybackslash}m{2.8cm}
}
\hline
\textbf{Model} 
& \textbf{Para. (M)} 
& \textbf{LR} 
& \textbf{LR sched.} 
& \textbf{Warmup} 
& \textbf{Grad. Acc.} 
& \textbf{Batch size} \\
\hline
Whisper-Medium-Wu 
& 769 
& $8 \times 10^{-5}$ 
& WarmupLR 
& 4000 steps 
& 4 
& Dynamic (24k frames) \\

Conformer-U2pp-Wu 
& 123 
& $1 \times 10^{-3}$ 
& WarmupLR 
& 25000 steps 
& 4 
& Dynamic (60k frames) \\
Step-Audio2-Wu-ASR 
& 7000 
& $1 \times 10^{-5}$ 
& WarmupLR 
& 0.05 ratio 
& 8 
& 8 \\
CosyVoice2-Wu-CPT
& 500 
& $1 \times 10^{-4}$ 
& WarmupLR 
& 25000 steps 
& 2 
& Dynamic (10k frames) \\
CosyVoice2-Wu-SFT
& 500
& $1 \times 10^{-5}$ 
& ConstantLR
& 0
& 2 
& Dynamic (2k frames) \\
CosyVoice2-Wu-SS
& 500
& $5 \times 10^{-6}$ 
& ConstantLR
& 0
& 2
& Dynamic (1k frames) \\
CosyVoice2-Wu-instruct
& 500
& $5 \times 10^{-6}$ 
& ConstantLR
& 0
& 2
& Dynamic (1k frames) \\
\hline
\end{tabular}
\end{table*}

\subsection{Details of Experiments}
\textbf{Optimization and Training Hyperparameters.}
The training hyperparameters for different models are summarized in Table~\ref{tab:training_hyperparams}.

\begin{figure}[htbp]
\centering
\includegraphics[clip, trim=3.5cm 1.5cm 4cm 1cm, width=0.8\linewidth]{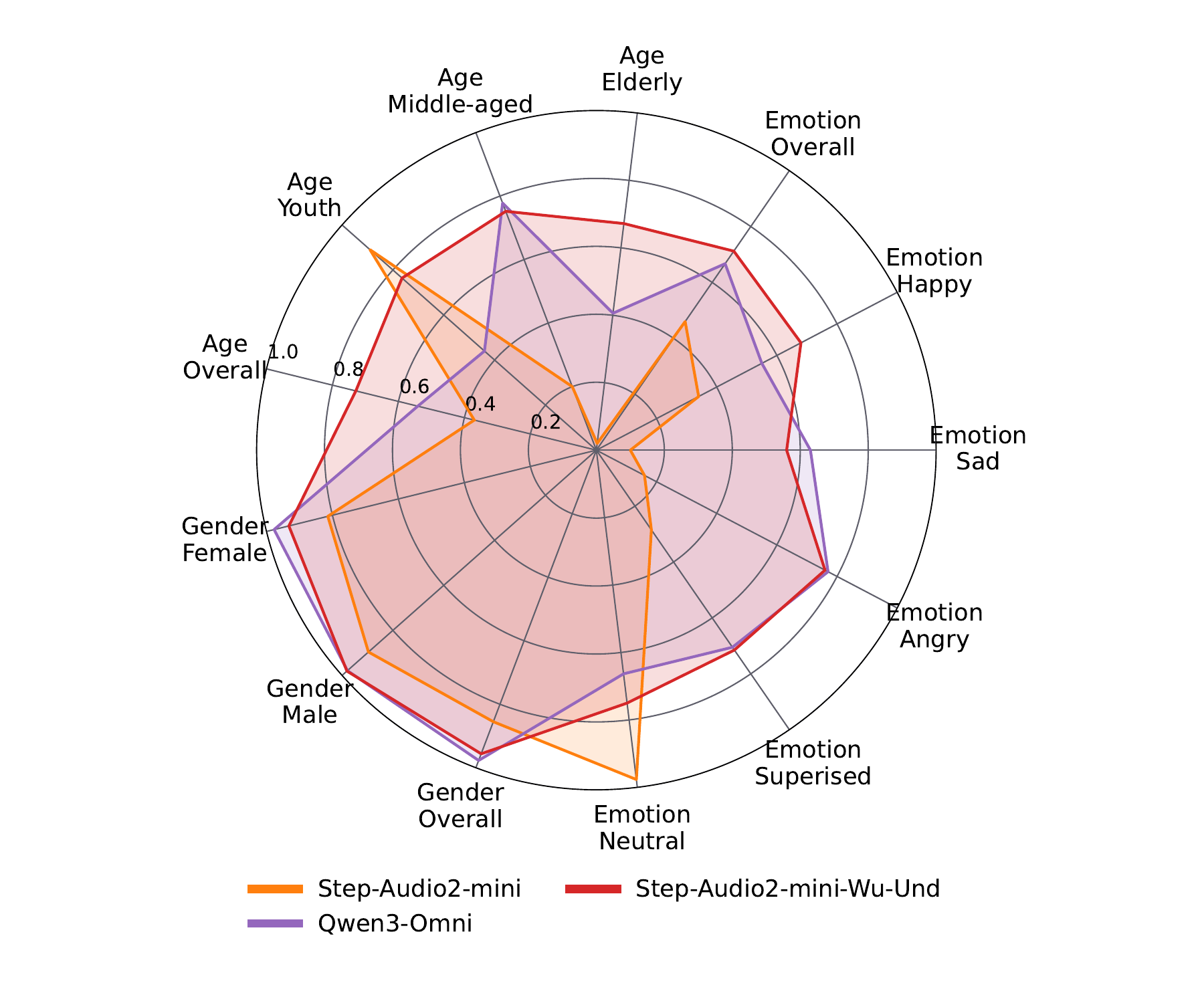}
\caption{Supplementary comparison results on age, gender, and emotion recognition tasks.}
\label{fig:supp_results}
\end{figure}

\textbf{Supplementary Experimental Results.}
To provide a more detailed analysis of model performance on WS-Wu-Bench, we report category-wise classification accuracy for gender, age, and emotion across WS-Wu-Und and baseline models.
As shown in Figure~\ref{fig:supp_results}, our model achieves more balanced and consistently higher performance across different paralinguistic categories, demonstrating stronger paralinguistic recognition capability.

\textbf{Prompt Templates Used in Experiments.}
For all speech understanding tasks, the same prompt templates are used during both training and inference to avoid potential distribution mismatch.For instruct TTS tasks, instruction-based templates are likewise employed in both training and inference to enable explicit instruction control.
Table~\ref{tab:prompts} lists the prompt formulations for each task.
The English prompts are shown for ease of presentation, while equivalent instructions are employed in other languages when applicable.
\begin{table}[htbp]
    \centering
    \renewcommand{\arraystretch}{1.4}
    \footnotesize
    \caption{Details of prompt templates used in speech understanding and instruct TTS experiments.}
    \label{tab:prompts}
    \begin{tabular}{
        >{\centering\arraybackslash}m{1cm}
        >{\centering\arraybackslash}m{5.6cm}
    }
        \toprule
        \textbf{Task Type} & \textbf{Prompt Content} \\
        \midrule
        ASR
        & Please transcribe the speech. \\
        \midrule
        AST
        & Please listen to this speech carefully and translate its content into Mandarin. \\
        \midrule
        Age
        & Based on the acoustic features of the speech, determine the speaker's age. Choose one label from: youth, middle-aged, or elderly. \\
        \midrule
        Gender
        & Based on the acoustic features of the speech, determine the speaker's gender. Choose one label from: male or female. \\
        \midrule
        Emotion
        & Based on the acoustic features and semantics of the speech, determine the emotion. Choose one label from: neutral, happy, sad, surprised, or angry. \\
        \midrule
        \makecell{Inst \\ Emo}
        & You must speak with anger, sadness, happiness, or surprise. \\
        \midrule
        \makecell{Inst \\ Pro}
        & This is a man or woman speaking in a low or high voice and at a slow or fast pace. \\
        \bottomrule
    \end{tabular}
\label{tab:prompts}
\end{table}


\end{document}